\documentclass[11pt,nofootinbib,aps,preprint]{revtex4-1}

\usepackage{amsmath,mathrsfs}
\usepackage{graphicx,epsfig}
\usepackage{latexsym,amssymb}
\usepackage{epsf}
\usepackage{ifpdf,natbib,lineno}

\def\sech{\hspace{0.75mm}{\rm sech}}
\usepackage{xcolor}

\begin{document}

\title{{\bf  The Affective Factors on the Uncertainty in the Collisions  of  the Soliton Solutions of the Double Field sine-Gordon System}}

\author{ M. Mohammadi$^{1}$} \email{physmohammadi@pgu.ac.ir}   \author{N. Riazi$^{2}$}\email{n_riazi@sbu.ac.ir} \address{$^1$Physics Department, Persian Gulf University, Bushehr 75169, Iran. \\ $^2$Department of Physics, Shahid Beheshti University, Evin, Tehran 19839, Iran.}

\date{\today}

\begin{abstract}
Inspired by the well known sine-Gordon equation, we present a  symmetric coupled system of  two real scalar fields in $1+1$ dimensions. There are three different topological soliton solutions which be labelled according to  their topological charges. These solitons can absorb
some localized non-dispersive wave packets in collision processes.
It will be  shown numerically,  during  collisions between solitons,  there will be an uncertainty which originates from  the amount of the maximum amplitude and arbitrary initial phases of the trapped wave packets.
\end{abstract}

\maketitle

 \textbf{Keywords} : { sine-Gordon, coupling, nonlinear, soliton, wave packet, uncertainty.}

\section{Introduction}

Relativistic solitons, including those of the conventional sine-Gordon (SG) equation, exhibit
remarkable similarities with classical particles. They exert short range forces on each other
and make collisions, without losing their identities \cite{rajarama,Das,lamb,Drazin}. They are localized objects and do
not disperse while propagating in the medium. The integrable SG  system has been considered in recent investigations, it has various  applications in many branches of physics \cite{Riazi1,a11,a12,a13,a14,a15}.
Because of their wave nature, they do tunnel
a barrier in certain cases, although this tunnelling is different from the well-known quantum
version \cite{Riazi1,Riazi3}. Topological solitons are stable, due to the boundary conditions at spatial infinity.
Their existence, therefore, is essentially dependent on the presence of degenerate vacua \cite{Lee}.

Topology provides an elegant way of classifying solitons in various sectors according to
the mappings between the degenerate vacua of the field and the points at spatial infinity.
For the sine-Gordon system in $1+1$ dimensions, these mappings are between $\phi =n\pi$, $n\in Z$
and $x = \pm \infty$, which correspond to kinks and anti-kinks of the SG system.

 Coupled systems of scalar fields have been investigated by many authors \cite{rajarama,Bazeian,Bazeia,Riazi1,Riazi4}.
Bazeia et al. \cite{Bazeia} considered a system of two coupled real scalar fields with a particular self-interaction
potential such that the static solutions are derivable from first order coupled differential
equations. Riazi et al. \cite{Riazi4} employed the same method to investigate the stability of the
single soliton solutions of a particular system of this type. Moreover, they used SG system to make a coupled  system of two real scalar fields with a rich structure and dynamics \cite{Riazi1}.

Inspired  by the well-known
properties of the SG equation, we will introduce a symmetric coupled systems of  two real scalar fields.
It can be considered as two  similar SG equations which are coupled to each other. It has  three types of soliton solutions which are named $H$ (horizontal), $V$ (vertical) and $D$ (diagonal) solitons. $H$ and $V$ solutions are nothing but the  usual SG kink (anti-kink) solutions.  They can combine to other and form a $D$-soliton. Moreover, it will be shown numerically and analytically  that a $H$ ($V$)-soliton can absorb  small wave packets which evolve according to a Schr\"{o}dinger like equation. These small wave packets do not make any considerable change in the particle aspects of a soliton but they have a main role in the collisions. These packets  lead to an uncertainty in the collision processes related to  trivial  initial phases.   In fact, initial phases behave  like  hidden variables  which lead to  different fates in the collision processes.   Moreover,  these wave-packets do not disperse and they satisfy a similar deBroglie's wavelength-momentum relation \cite{Riazi}.

Note that the term ``soliton'' is used throughout this paper for
localized solutions. The problem of integrability of the model is not addressed here. Such
non-dispersive solutions are called ``lumps'' by Coleman \cite{Coleman} to avoid confusion with true
solitons of integrable models. However, it has now become popular to use the term soliton
in its general sense.

The organization of this paper is as follows: In section  \ref{sec2}, we introduce the lagrangian
density, dynamical equations, and conserved currents of the proposed model.  In section
\ref{sec3}, some exact and numerical solutions, together with the corresponding topological charges and energies are derived.
The necessary nomenclature and the other  properties  of the  single soliton solutions   are also introduced in this section.  In section \ref{sec5}, we proceed  analytically to understand why a soliton can absorb a wave packet, and will provide a detailed discussion about the  internal modes of each single soliton solution. In section \ref{sec6},  we  will see numerically an uncertainty in collision processes which originates from the wave aspect of  the soliton solutions.

\section{DYNAMICAL EQUATIONS, CONSERVED CURRENTS, AND TOPOLOGICAL CHARGES}\label{sec2}

Within a relativistic formulation, the well known sine-Gordon (SG) equation for a real field ($\phi$) in 1+1 space-time dimensions can be written as
\begin{equation} \label{SG}
\Box \phi =-\sin(2\phi).
\end{equation}
The factor $2$ in the argument of $\sin(2\phi)$ dose not change the main properties of the solutions and we have included it for convergence.
This equation can be extracted from the following Lagrangian density:
\begin{equation} \label{Lag}
 {\cal L}=\frac{1}{2}\partial_{\mu}\phi\partial^{\mu}\phi-\sin^{2}(\phi).
\end{equation}
The special traveling wave  solutions of this equation are topological objects which are called kinks and antikinks:
\begin{equation} \label{sdl}
\phi_{v}(x,t)=\varphi(\gamma(x-vt))=2\tan^{-1}[\exp(\pm \sqrt{2}\gamma (x-vt))],
\end{equation}
where $v$ is the velocity, $\gamma=1/\sqrt{1-v^2}$ and $+$ ($-$) is denoted   for kinks (antikinks).  For any kink (antikink) solution (\ref{sdl}), the related localized energy density function is
\begin{equation} \label{sd11}
\varepsilon(x,t)=2\sech^{2}[\sqrt{2}\gamma (x-vt)].
\end{equation}
 Such solutions (\ref{sdl}) are very stable and without any deformation reappear after collisions. In this respect, they can be imagined as non-interacting real particles in 1+1 space-time.

\begin{figure}[ht!]
 \centering
 \includegraphics[width=130mm]{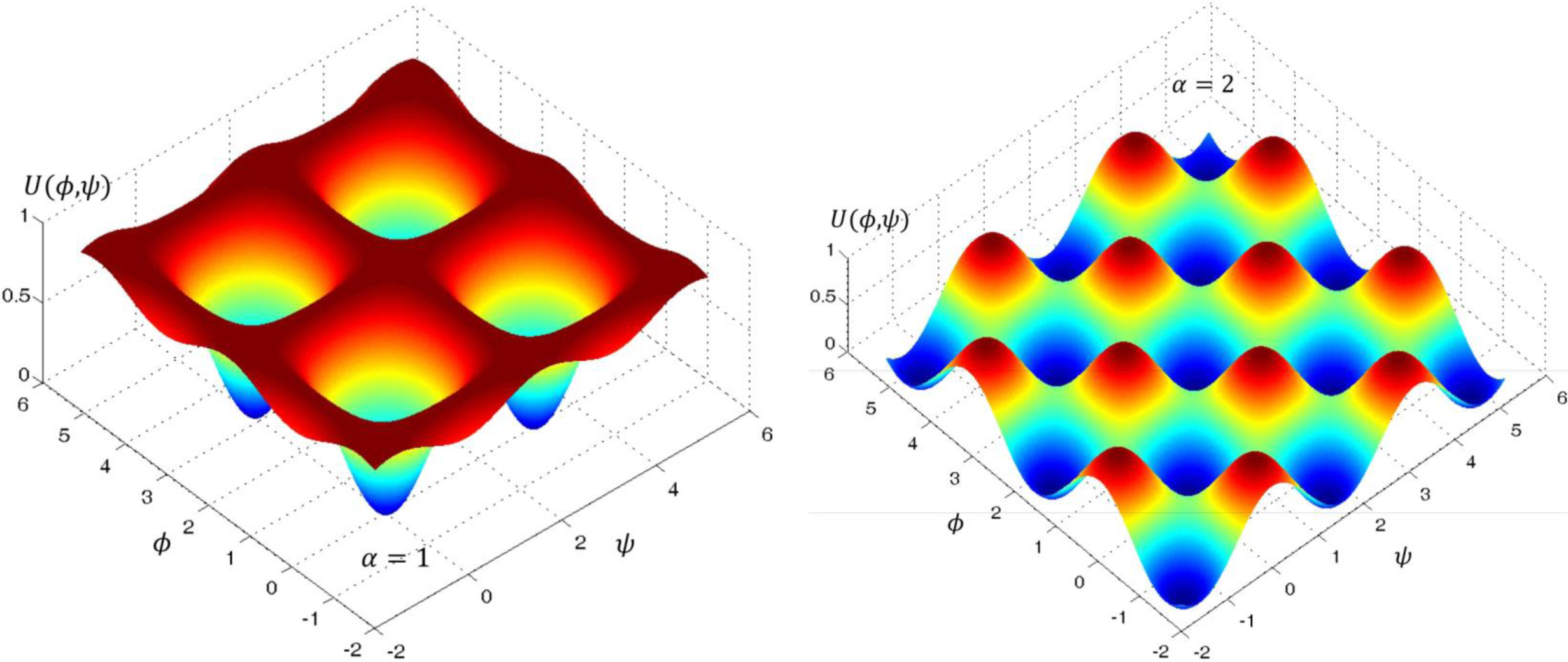}
 \caption{The 3D shape of the potential (\ref{e2}) for  two special cases $\alpha=1$  and $\alpha=2$. It obvious that for the case $\alpha=1$, as  a representative of the systems  $\alpha<2$, the vacua are $(\phi,\psi)=(N_{1}\pi,N_{2}\pi)$, where $N_{1}$ and $N_{2}$ are any arbitrary integer number.} \label{3D}
 \end{figure}

We employ the SG system  to construct  a new system in 1+1  space-time dimensions with two real scalar fields $\phi$ and $\psi$,  by introducing  a  new Lagrangian density:
\begin{equation} \label{e1}
 {\cal L}=\frac{1}{2}\partial_{\mu}\phi\partial^{\mu}\phi+\frac{1}{2}\partial_{\mu}\psi\partial^{\mu}\psi
-U(\phi,\psi),
 \end{equation}
 where
 \begin{equation} \label{e2}
U(\phi,\psi)=\sin^{2}(\phi)+\sin^{2}(\psi)-\alpha\sin^{2}(\phi)\sin^{2}(\psi),
  \end{equation}
is the  field potential and $ \alpha $ is  a real   constant parameter. The field potentials
for two values of $\alpha= 1$ and $\alpha= 2$ are shown in Fig.~\ref{3D}.   It is obvious that $\alpha$ controls the strength of interactions between  fields  $\phi $ and $ \psi $.  For example, if $ \alpha=0 $, there is no interaction between fields and we have two independent sine-Gordon (SG) field  equations for $ \phi $ and $ \psi $.
For low energy excitations ($\phi,\psi \ll 1$), the potential for the coupled fields can be expanded up to the fourth order:
\begin{equation} \label{4payeh}
U(\phi,\psi)\approx \phi^{2}+\psi^{2}-\alpha\phi^{2}\psi^{2}.
\end{equation}
According to the standard quantum field theory, this leads to two equal  mass terms $\phi^{2}$ and $\psi^{2}$, and a four legs interaction term $\alpha\phi^{2}\psi^{2}$.

 From the Lagrangian density (\ref{e1}), we obtain the following equations for $ \phi $ and $ \psi $, respectively
\begin{equation} \label{e3}
\Box \phi = -\sin(2\phi)(1-\alpha\sin^2(\psi)),
\end{equation}
 and
 \begin{equation} \label{e4}
 \Box \psi = -\sin(2\psi)(1-\alpha\sin^2(\phi)).
 \end{equation}
 Since the Lagrangian density (\ref{e1}) is Lorentz invariant, the corresponding energy-momentum
 tensor \cite{rajarama} is
 \begin{equation} \label{e5}
 T_{\mu\nu}=\partial_{\mu}\phi\partial_{\nu}\phi+\partial_{\mu}\psi\partial_{\nu}\psi-\eta_{\mu\nu}{\cal L},
  \end{equation}
 which satisfies the conservation law:
\begin{equation} \label{e6}
\partial_{\mu}T^{\mu\nu}=0.
\end{equation}
 In equation (\ref{e5}), $\eta_{\mu\nu}= \textrm{diag}(1,-1)$ is the metric of the 1 + 1 dimensional Minkowski space-time.    The Hamiltonian (energy) density is obtained from Eq.~(\ref{e5}) according to
 \begin{equation} \label{e7}
 {\cal H}=\varepsilon(x,t)=
 T^{00}=\frac{1}{2}\dot{\phi}\dot{\phi}+\frac{1}{2}\phi^{'}\phi^{'}+\frac{1}{2}\dot{\psi}\dot{\psi}+\frac{1}{2}\psi^{'}\psi^{'}+U(\phi,\psi),
 \end{equation}
 in which dot and prime denote differentiation with respect to $ t $ and $ x $, respectively. Note that we have assumed  c=1 throughout   this paper.
Moreover, the following topological currents can be defined:
\begin{equation} \label{e8}
J_{V}^{\mu}=C_{1}\epsilon^{\mu\nu}\partial_{\nu}\phi, \quad J_{H}^{\mu}=C_{2}\epsilon^{\mu\nu}\partial_{\nu}\psi,
\end{equation}
 where $ \epsilon^{\mu\nu} $ is the completely antisymmetric tensor and $ C_{1} $ and $ C_{2} $ are arbitrary constants.
 The subscripts $ H $, $ V $ and $ D $, denote “horizontal”, “vertical” and “diagonal” which will be explained later. These currents ($ J_{H,V}^{\mu}$) are conserved independently:
 \begin{equation} \label{e9}
\partial_{\mu}J_{V}^{\mu}=0, \quad \partial_{\mu}J_{H}^{\mu}=0,
 \end{equation}
and lead to quantized constant topological charges. The corresponding topological charges are given by:
\begin{equation} \label{e10}
 Q_{V}=\int_{-\infty}^{+\infty} J_{V}^{0}dx=C_{1}[\phi(+\infty)-\phi(-\infty)],
\end{equation}
 \begin{equation} \label{e11}
  Q_{H}=\int_{-\infty}^{+\infty} J_{H}^{0}dx=C_{2}[\psi(+\infty)-\psi(-\infty)].
 \end{equation}
In this paper,  we set $C_{1}=C_{2}=\frac{1}{\pi}$, for convenience.

 \section{SINGLE-SOLITON SOLUTIONS \label{sec3}}

\begin{figure}[ht!]
 \centering
 \includegraphics[width=160mm]{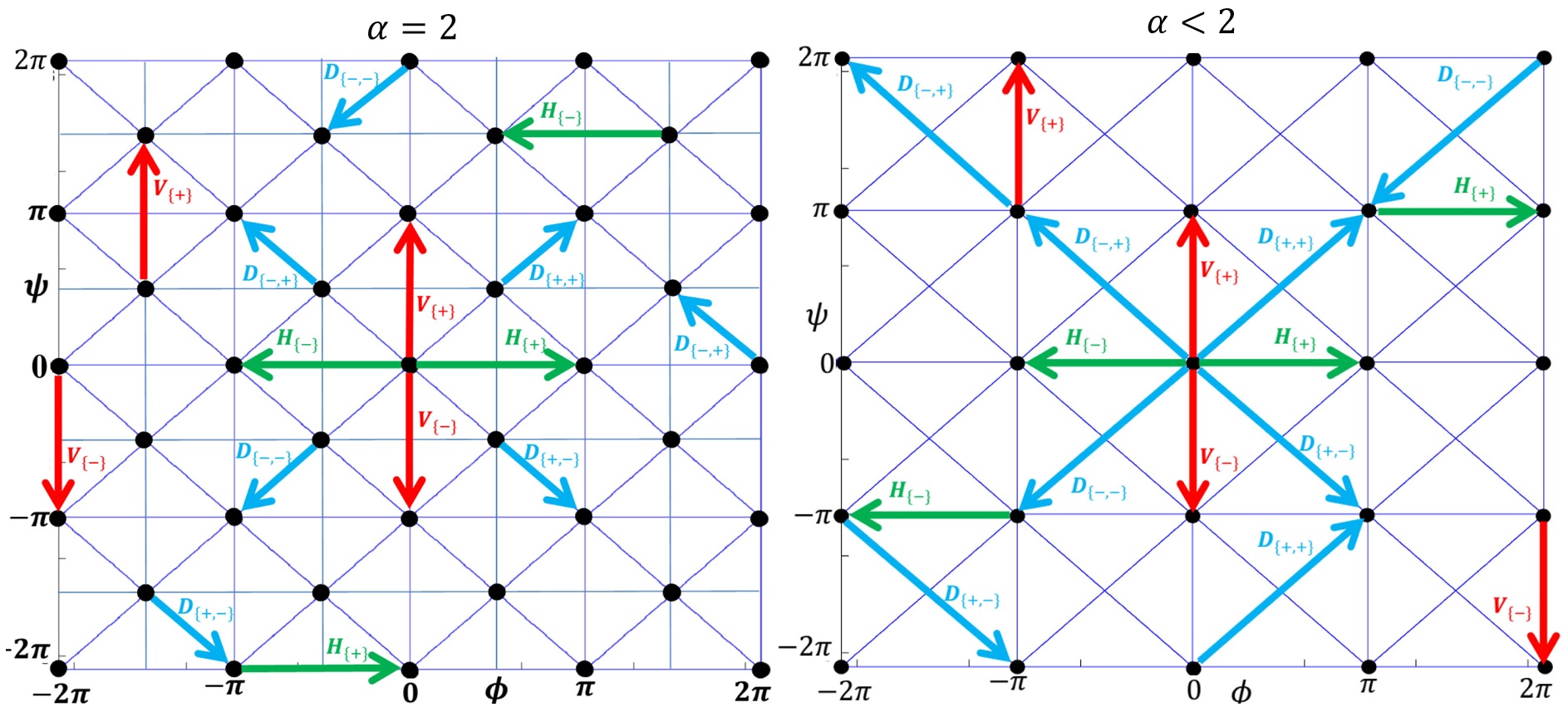}
 \caption{According to Fig.~\ref{3D}, if one considers just the vacuum points, a mesh of the regular points is obtained on the $(\phi,\psi)$ plane. Black points are correspond to vacua. Nomenclature of horizontal (H), vertical (V) and diagonal (D) solutions according to the boundary conditions on the $(\phi,\psi)$ plane. Each line which connects two neighbouring vacua  can be correspond to a stationery soliton solution. It is easy to understand there are infinite soliton solutions. Note that the subscripts $\{+\}$ ($\{-\}$) is used for a kink (antitank) solution.} \label{vacua}
 \end{figure}

 Each static stable  soliton solution should  vary between two neighbouring  vacuum points  and its  energy must be minimum  among infinite ways which can connect  the two vacua,  i.e. they must be stationary. Namely, if we consider the systems which were  introduced in the previous section, it is easy to show that there are three kinds of static soliton solutions:   $H$ (horizontal), $V$ (vertical) and $D$ (diagonal) types, depending on the beginning and ending points of the solution on the $(\phi,\psi)$ plane \cite{Riazi1}. They are shown symbolically in  Fig.~\ref{vacua}.

\begin{figure}[ht!]
  \centering
  \includegraphics[width=120mm]{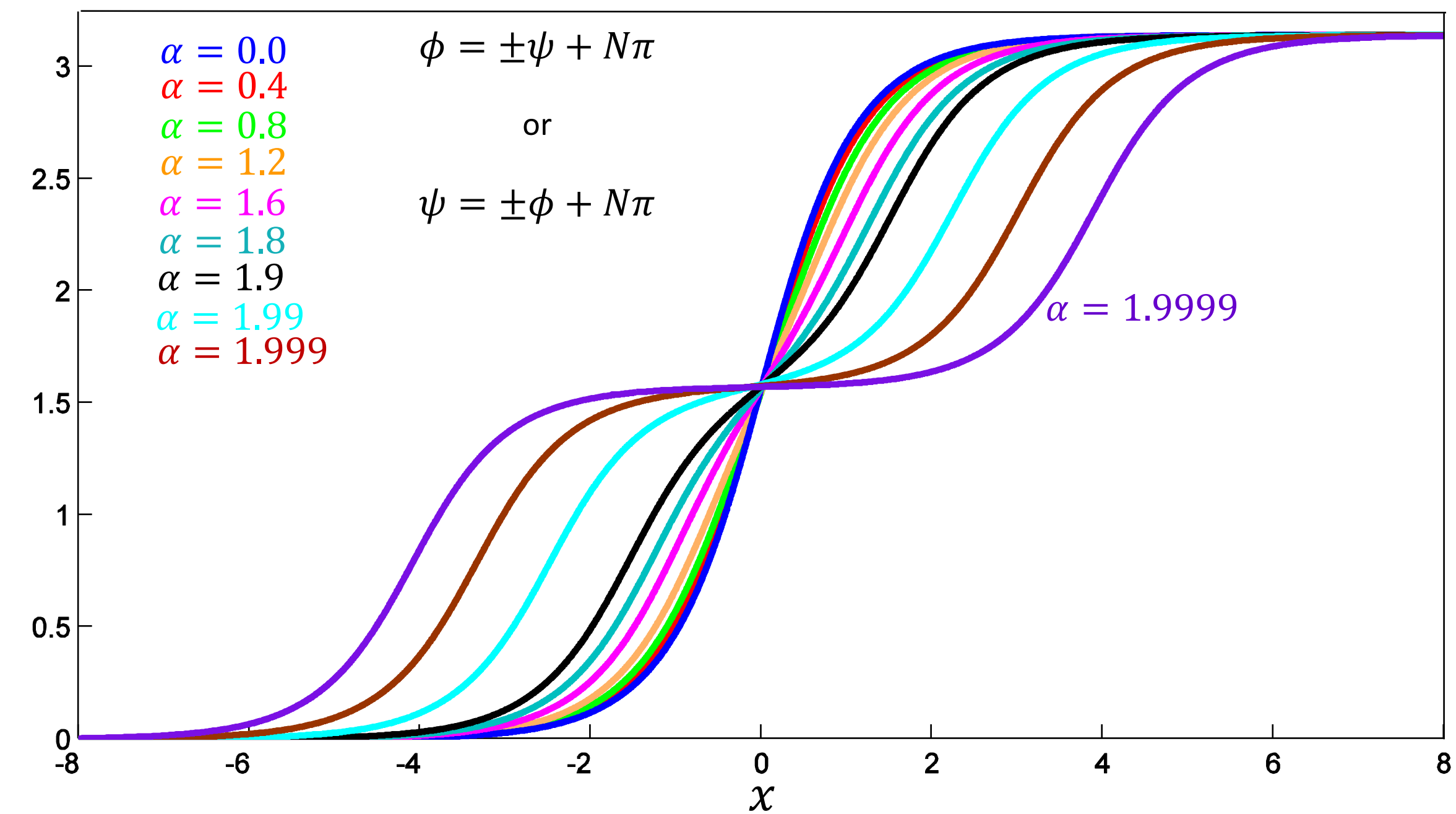}
  \caption{For different $\alpha$'s, there are different $D$-soliton solutions.} \label{Dsolitons}
  \end{figure}

\begin{figure}[ht!]
 \centering
 \includegraphics[width=130mm]{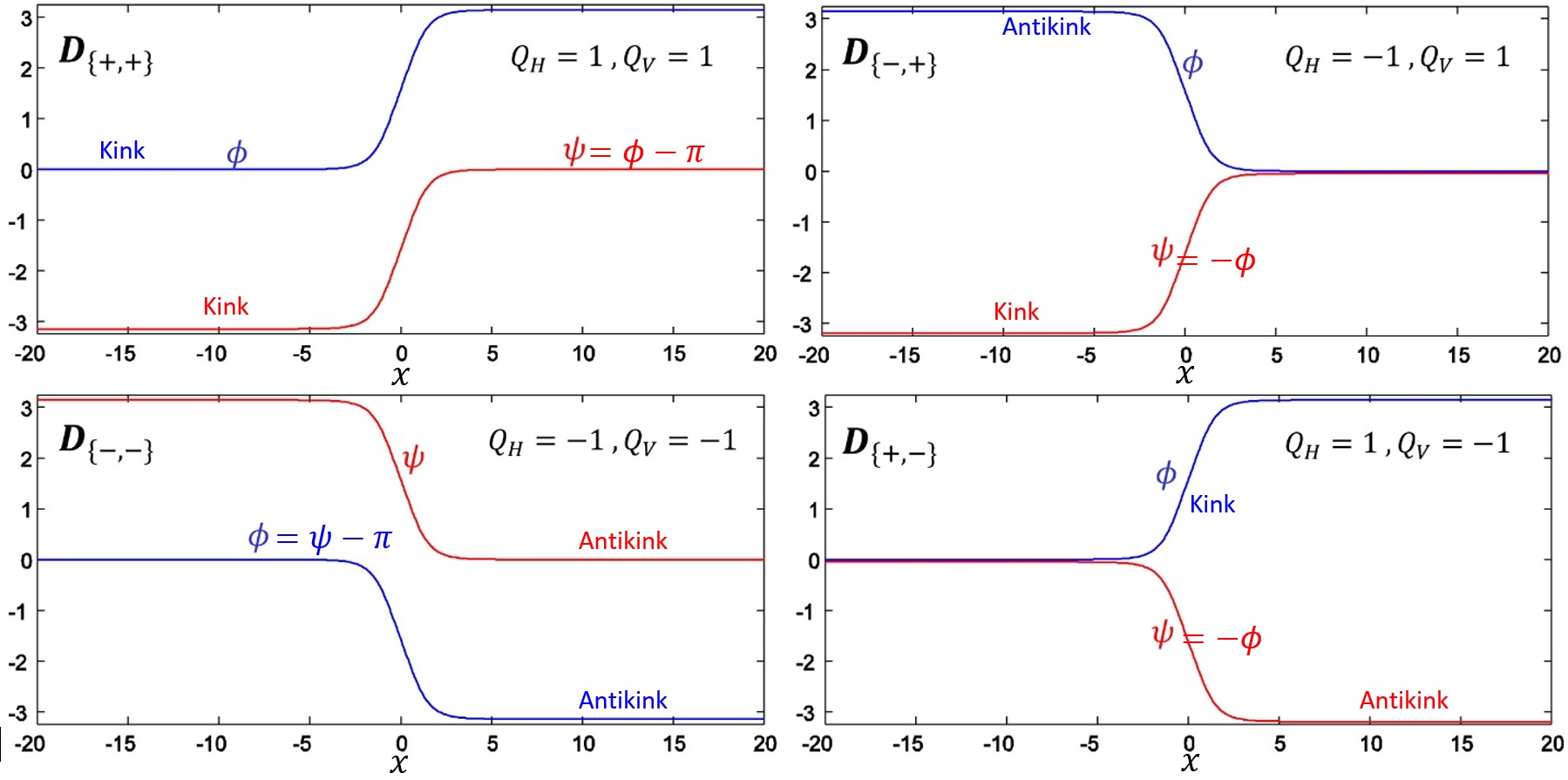}
 \caption{There are  four types of D-soliton solutions for $\alpha=1$ (as a representative of the systems with $\alpha<2$) which are characterized by different permutations  of $Q_{H}$ and $Q_{V}$ signs. In other words, they are characterized to four different  combinations of the kinks and antikinks related  to  fields $\phi$ and $\psi$.} \label{ad}
 \end{figure}

In general, there are two scalar fields $\phi$ and $\psi$ which are coupled together via  the dynamical equations (\ref{e3}) and (\ref{e4}). For three special situations, these coupled equations reduce  to simple formats. First, if one sets  $\psi=N_{2}\pi$ ($N_{2}=0,\pm 1,\pm 2, \cdots$), Eq.~(\ref{e4}) is satisfied automatically and Eq.~(\ref{e3}) turns to the same original SG equation (\ref{SG}) with the same well-known kink ($H_{\{+\}}$) and antikink ($H_{\{-\}}$) solutions (\ref{sdl}). Second, on the contrary, for a fixed value of  $\phi=N_{1}\pi$ ($N_{1}=0,\pm 1,\pm 2, \cdots$), Eq.~(\ref{e3}) is satisfied automatically and  Eq.~(\ref{e4}) leads to the same kink ($V_{\{+\}}$) and antikink ($V_{\{-\}}$) solutions (\ref{sdl}) as well. Third, if one considers the situations for which $\phi=\pm\psi+N_{1}\pi$ or  $\psi=\pm\phi+N_{2}\pi$ (i.e. $D$-solitons), Eqs.~(\ref{e3}) and (\ref{e4}) both  turn  to an identical  dynamical equation as follows:
\begin{equation} \label{e31}
\Box \phi = -\sin(2\phi)(1-\alpha\sin^2(\phi))=-\dfrac{d{\cal U}}{d\phi},
\end{equation}
where
\begin{equation} \label{e32}
{\cal U}(\phi)=\sin^{2}(\phi)-\frac{\alpha}{2}\sin^{4}(\phi),
\end{equation}
is  considered as a new field potential just for $D$-solitons.  This potential (\ref{e32}) for $\alpha\leq 2$ is always positive and its vacua are  $\phi=N\pi$ ($N=0,\pm 1,\pm 2, \cdots$). It again, depending on $\alpha$,  leads to new types  of the kink and antikink  solutions which  unfortunately  we could  not find their explicit forms at all.  However,   by a Runge-Kutta  method,  one can simply obtain them  numerically (see Figs.~\ref{Dsolitons} and \ref{ad}).
 Thus, the evolution of   such solutions ($D$-solitons) can be simulated by a single  non-linear  field equation (\ref{e31}).
Fortunately for case $\alpha=2$,  the PDE  (\ref{e31})  reduces  to a  new version of the SG equation:
 \begin{equation} \label{e13}
 \Box \phi = -\frac{1}{2}\sin(4\phi)=-\dfrac{d}{d\phi}[\frac{1}{4}\sin^{2}(2\phi)],
 \end{equation}
for which ${\cal U}(\phi)=\frac{1}{4}\sin^{2}(2\phi)$. The $D$-soliton solutions of this  equation (\ref{e13}) are presented  in Table.~\ref{alpha2}. The rest energy in this case can be calculated easily equal to  $E_{o}=\sqrt{2}$ which is exactly half of  $H$ and $V$  rest energy.

Briefly, there are two different types of $H$ ($V$)-solitons which are denoted by $H_{\{+\}}$ ($V_{\{+\}}$) and $H_{\{-\}}$ ($V_{\{-\}}$). The subscript $\{+\}$ ($\{-\}$)  is refereed to a kink (anti-kink) with positive (negative) topological charge. All exact static $H$ and $V$-solitons and their features are summarized in Table.~\ref{alpha1}. If a kink or anti-kink of $\phi$ field lives with another kink or anti-kink of $\psi$ field, we have a $D$-soliton. There are 4 types of $D$-solitons which correspond to different combinations
  of kink or anti-kink  for $\phi$ and $\psi$. Namely, a $D_{\{+,-\}}$-soliton  is composed of a kink with $Q_{H}>0$ for the $\phi$ field beside an anti-kink with $Q_{V}<0$ for the $\psi$ field. In an equivalent statement, the first and second subscript in a $D$-soliton is used to characterize $Q_{H}$ and $Q_{V}$ signs respectively (see Fig.~\ref{ad}).

\begin{figure}[ht!]
  \centering
  \includegraphics[width=140mm]{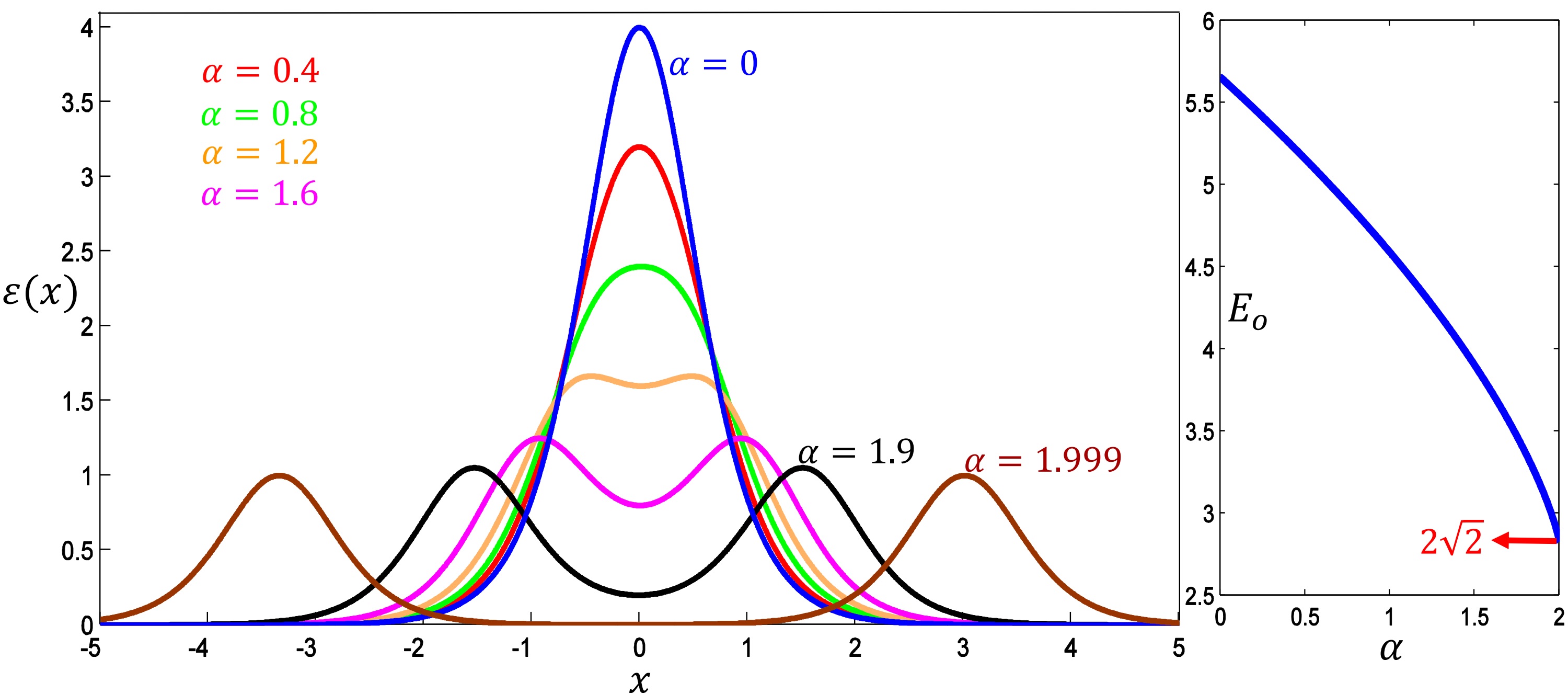}
  \caption{The curves in the left hand side  are  the   energy density functions  for  different   non-moving D-solitons. The right hand side  is  the rest energy of  the $D$-solitons versus $\alpha$.} \label{DE}
  \end{figure}

 \begin{table}
 \begin{center}
 \begin{tabular}{ | l |p{11cm}| l | l | l |}

     \hline
     \centering{Type} & \centering{Solution} & $Q_{H}$ & $Q_{V}$ &\quad $E_{o}$\\ \hline
     \quad $H_{\{\pm\}}$ & \centering{$\phi=2\tan^{-1}[\exp(\pm\sqrt{2}x)]+N_{1}\pi,\quad\quad \psi=N_{2}\pi$} & \centering{$\pm 1$} & $0$ &\quad $2\sqrt{2}$ \\[0.5cm]
     \quad $V_{\{\pm\}}$ & \centering{$\phi=N_{1}\pi,\quad\quad \psi=2\tan^{-1}[\exp(\pm\sqrt{2}x)]+N_{2}\pi$} & $0$ & $\pm 1$ &\quad $2\sqrt{2}$ \\    [0.5cm]
     \hline
    \end{tabular}
 \caption{Exact static soliton solutions of the systems $\alpha<2$ and the corresponding horizontal and vertical topological charges and rest energies. $N_{1}$ and $N_{2}$ are just arbitrary integer constants.}\label{alpha1}
 \end{center}
 \end{table}

 \begin{table}
  \begin{center}
  \begin{tabular}{ | l |p{8cm}| l | l | l |}

      \hline
      \centering{Type} & \centering{Solution} & $Q_{H}$ & $Q_{V}$ & $E_{0}$\\ \hline
      \quad $H_{\{\pm\}}$ & \centering{$\phi=2\tan^{-1}[\exp(\pm\sqrt{2}x)]+\dfrac{N_{1}\pi}{2},\quad \psi=\dfrac{2N_{2}+N_{1}}{2}\pi$} & \centering{$\pm 1$} & $0$& $2\sqrt{2}$ \\[0.5cm]
      \quad $V_{\{\pm\}}$ & \centering{$\phi=\dfrac{2N_{1}+N_{2}}{2}\pi,\quad \psi=2\tan^{-1}[\exp(\pm\sqrt{2} x)]+\dfrac{N_{2}\pi}{2}$} & $0$ & $\pm 1$ & $2\sqrt{2}$\\  [0.5cm]
      \quad $D_{\{\pm,\pm\}}$ & \centering{$\phi=\pm\psi+N_{2}\pi=\tan^{-1}[\exp(\pm\sqrt{2}x)]+\dfrac{N_{1}\pi}{2}$} & $\pm \frac{1}{2}$ & $\pm \frac{1}{2}$ & $\sqrt{2}$\\  [0.5cm]
      \hline
     \end{tabular}
  \caption{Exact static soliton solutions and the corresponding horizontal and vertical topological charges and rest energies for case $\alpha=2$ . $N_{1}$ and $N_{2}$ are just arbitrary integer constants.}\label{alpha2}
  \end{center}
  \end{table}

The rest energy  of a  soliton solutions is obtained by   integration  of the static kink energy density
 \begin{equation} \label{e12}
 E_{o}=\int_{-\infty}^{+\infty}\varepsilon(x) dx.
 \end{equation}
 The rest energy  of $H$ and $V$-soliton solutions for  systems with   $\alpha\leq 2$ all are equal to $E_{o}=2\sqrt{2}$.  But for different $D$-solitons which are identified with different $\alpha$'s (Fig.~\ref{Dsolitons}), there are different energy density functions and rest energies (see  Fig.~\ref{DE}).

 From now, we only consider systems with a genuine interaction between their components ($\phi$ and $\psi$ fields), i.e. those for which $\alpha<2$ (except $\alpha=0$). All systems which are  within
 this range have identical vacua and the right  part of Fig.~\ref{vacua} ($\alpha=1$)  can be generalized for all of them.

\section{INTERNAL MODES}\label{sec5}

\begin{figure}[ht!]
  \centering
  \includegraphics[width=140mm]{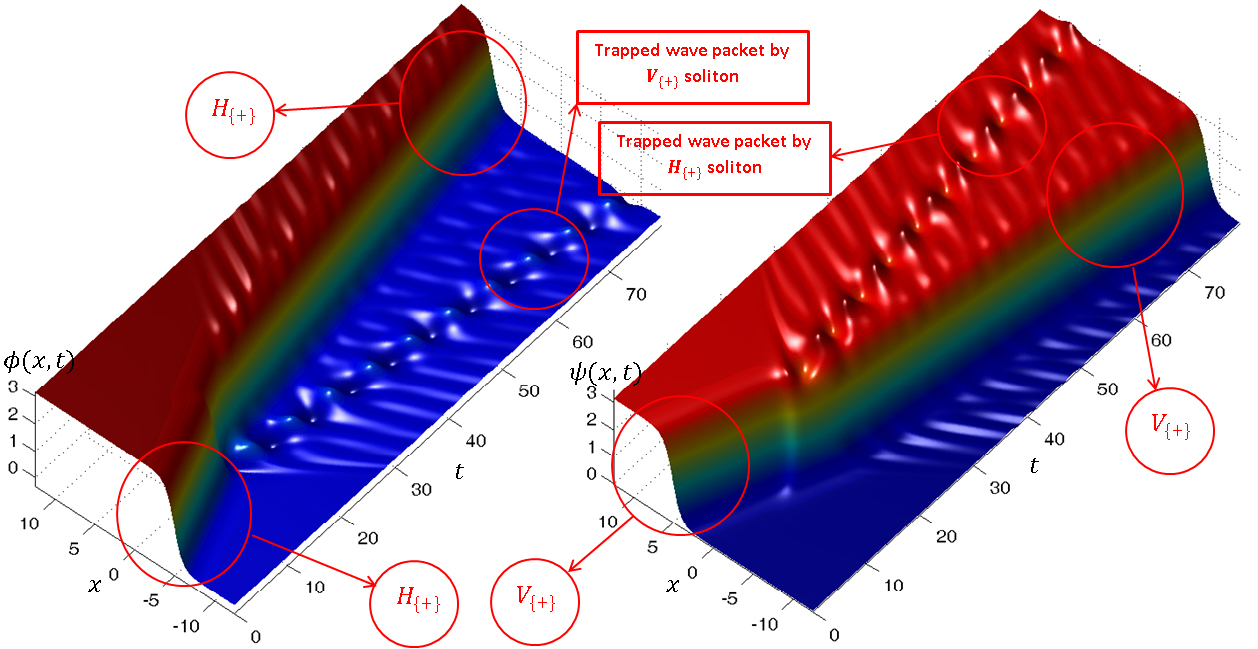}
  \caption{ A $V$ ($H$)-soliton can trap a wave packet from the  $\phi$ ($\psi$) field  for systems with  $0<\alpha<2$. This special Fig is the field representation of a $H_{\{+\}}-V_{\{+\}}$  collision for $\alpha=1.2$ with initial speed $v=0.3$. Note that, the shape of the kinks themselves  stay  almost unchanged.    } \label{ghablphi}
  \end{figure}

In this section, analytically and numerically,  we showed that  $H$,  $V$ and $D$-solitons  can keep a constantly oscillating behaviour   for  $0<\alpha<2$. In  similar situations \cite{DC,G,Riazi4,OV,GH,MM1}, these  oscillations  have been   interpreted  as low energy trapped wave packets (see Fig.~\ref{ghablphi}). Numerically, it was seen  that  for systems with $\alpha<0$, there are no long lasting oscillations.
 To provide a quantitative discussion of this phenomenon we  follow the standard procedure and  apply a small oscillatory  perturbation to a non-moving $H$-soliton solution \cite{Bazeia, Riazi4}:
\begin{equation} \label{e21}
\phi_{o}(x,t)=\varphi(x)+\Phi(x)e^{-i(\Omega_{o} t-\beta)}+N_{1}\pi,\quad\quad \psi_{o}(x,t)=N_{2}\pi+\Psi (x)e^{-i(\omega_{o} t-\theta)}.
\end{equation}
in which $\varphi(x)=2\tan^{-1}[\exp(\pm\sqrt{2}x)]$ is the same exact non-moving soliton solution (\ref{sdl}), $\Phi$ (x) and $\Psi(x)$ are  small perturbations, $\theta$ and $\beta$  are arbitrary constant phases. Similarly, a little disturbed non-moving $V$-soliton can be shown as follows:
\begin{equation} \label{dfgt}
\phi_{o}(x,t)=N_{1}\pi+\Phi (x)e^{-i(\Omega_{o} t-\beta)},\quad\quad \psi_{o}(x,t)=\varphi(x)+\Psi(x)e^{-i(\omega_{o} t-\theta)}+N_{2}\pi.
\end{equation}
However,  inserting   the  little disturbed non-moving $H$-soliton (\ref{e21}), as an ansatz, into the non-linear differential equations (\ref{e3}) and (\ref{e4}) and expanding to the first order in $\Psi$ and $\Phi$, we obtain two independent eigenvalue equations for $\Psi$ and $\Phi$ respectively, which look like the Schr\"{o}dinger equation:
\begin{equation} \label{e22}
-\frac{d^{2}\Psi}{dx^{2}}+V_{1}(x)\Psi=\omega_{o}^{2}\Psi,
\end{equation}
and
\begin{equation} \label{e34}
-\frac{d^{2}\Phi}{dx^{2}}+V_{2}(x)\Phi=\Omega_{o}^{2}\Phi,
\end{equation}
where
\begin{equation} \label{e23}
   V_{1}(x)=2-2\alpha\sin^{2}(\varphi)=2-2\alpha\sech^{2}(\sqrt{2}x).
\end{equation}
and
\begin{equation} \label{e35}
   V_{2}(x)=\dfrac{d^2[\sin^{2}(\varphi)]}{d\varphi^2}=2-4\sech^{2}(\sqrt{2}x).
\end{equation}
It is easy to show that for the Schr\"{o}dinger-like  equation (\ref{e34}) with the well-known potential (\ref{e35}), there is  just a trivial solution  $\Phi(x)=\xi\dfrac{d\varphi}{dx}$ which corresponds to $\Omega_{o}=0$, and $\xi$ is  just any arbitrary  small number to be sure that $|\Phi|\ll 1$. In fact, this potential (\ref{e35}) is related to  the  well-known SG system  which were studied completely in Refs. \cite{GH,MM1}.  According to Eq.~(\ref{e21}), it is easy to show that   this trivial solution  (i.e. $\Phi(x)=\xi\dfrac{d\varphi}{dx}$ and $\Omega_{o}=0$) dose not have physical meaning and  is  just associated with an infinitesimal translation of the
static $H$-soliton:
\begin{equation} \label{fgj}
\phi_{o}(x)=\varphi(x)+\xi \frac{d\varphi_{o}}{dx}=\varphi (x+\xi).
\end{equation}
   In other words, for the Schr\"{o}dinger-like Eq.~(\ref{e34}) there is not any non-trivial solution corresponds to a channel to impose some permanent oscillations, hence  for  a  single $H$ ($V$)-soliton, $\Phi$ ($\Psi$) must be always   equal to zero  and the shape of the kinks and antikinks themselves  remain unchanged (see Fig.~{\ref{ghablphi}}).

\begin{figure}[ht!]
  \centering
  \includegraphics[width=130mm]{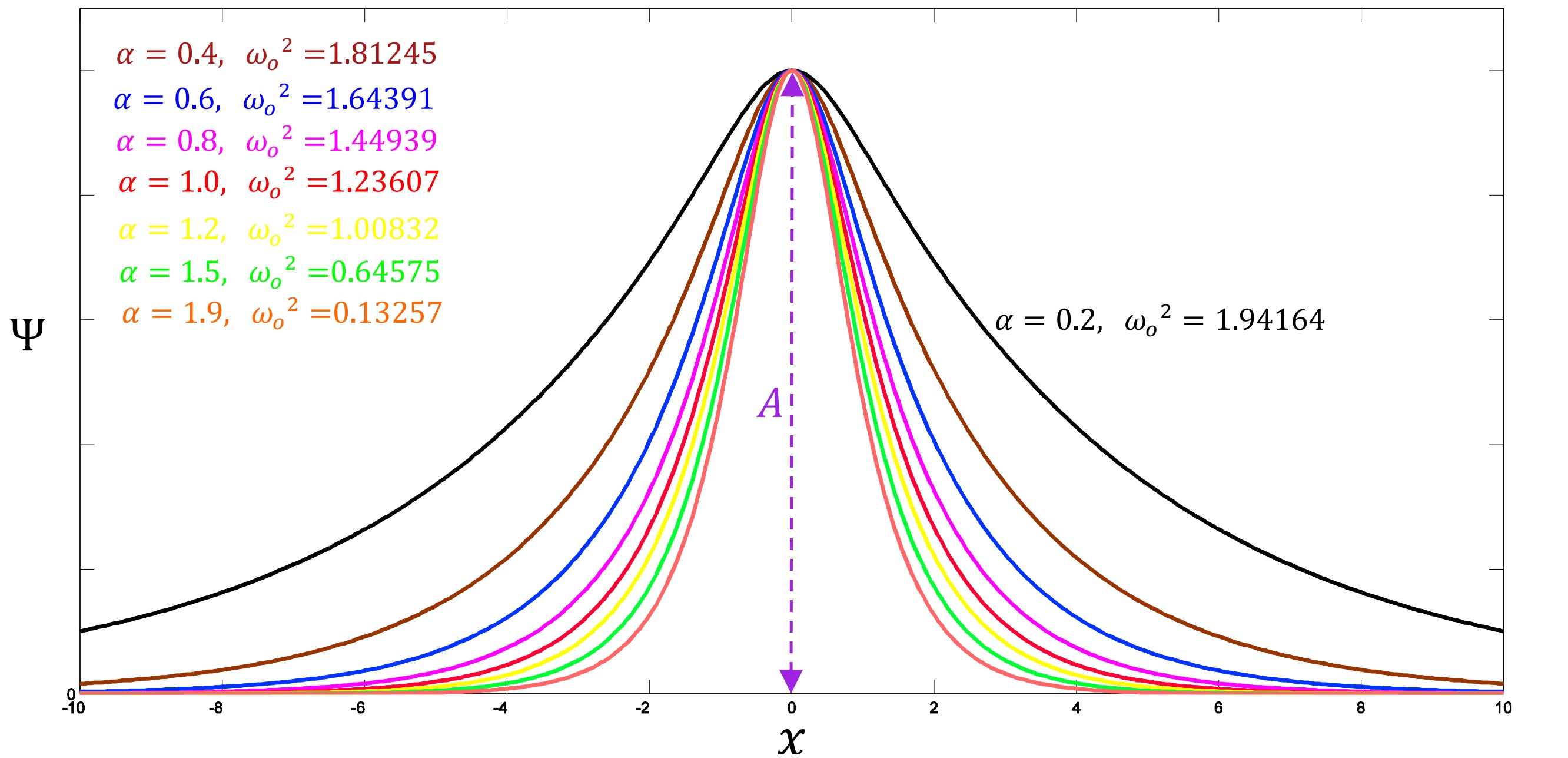}
  \caption{For any  arbitrary $\alpha$ ($0<\alpha<2$), the eigenvalue  equation (\ref{e22}) leads to a single bound state (eigenfunction) $\Psi$ with a special eigenvalue  $0<\omega_{o}^{2}<2$. Here, $A=\max(|\Psi|)$ is the maximum amplitude of the eigenfunction  $\Psi$ (trapped wave packet).   The existence of this bound state is the reason why we can see a permanent   oscillation after collisions in Fig.~\ref{ghablphi} for $H$-soliton. A similar reason exists for justifying the imposed   oscillation on the $V$-soliton in  Fig.~\ref{ghablphi}. } \label{bound}
 \end{figure}

Instead, the Schr\"{o}dinger-like equation (\ref{e22}) with the  potential (\ref{e23}) can be solved numerically, using a Runge-kutta method. It was shown that numerically for any arbitrary $\alpha$ ($0<\alpha<2$),  the Schr\"{o}dinger-like equation (\ref{e22}) leads to a single  bound state (internal mode) as a non-trivial localized solution with a special eigenvalue $0<\omega_{o}^{2}<2$ (see Fig.~\ref{bound}).
Therefore, for a  $H$ ($V$)-soliton there is a single non-trivial bound state $\Psi$ ($\Phi$) with $0<\omega_{o}^2<2$ ($0<\Omega_{o}^2<2$), but $\Phi=0$   ($\Psi=0$). In other words,   a  $H$ ($V$)-soliton
provides an attractive   potential  $V_{1}(x)$ which can trap some small localized waves of the adjacent field $\psi$ ($\phi$). The other cases for which  $\alpha<0$, we are faced with a potential barrier which cannot  have  any bound state.
Accordingly, one can  understand easily why for $H$ ($V$)-solitons  with $0<\alpha<2$, there is  a permanent oscillation after collisions,  because for  such solutions  there is an internal mode which can be considered as a channel for trapping  external energies and imposing some  permanent oscillations. Note that contrary to many double-field nonlinear systems, the two linear equations (\ref{e22}) and (\ref{e34}) are decoupled to first order in fields and linear perturbations of each field   do not see companion field  to first order.

 Note that the rest energy of a little  disturbed $H$ ($V$)-soliton, with a good approximation, is   equal to the same  undisturbed ones. In general, a little disturbed  non-moving $H$-soliton can be expressed as follows:
 \begin{equation} \label{vbg}
\phi_{o}(x,t)=\varphi(x)+\delta\phi(x,t)+N_{1}\pi,\quad\quad \psi_{o}(x,t)=N_{2}\pi+\delta\psi (x,t),
\end{equation}
where $\delta\phi(x,t)$ and $\delta\psi (x,t)$ can be considered  any arbitrary permissible  small deformations (variations). Note that, the ansatz    (\ref{e21})  is  a special kind of the general form  (\ref{vbg}).  If one inserts  theses deformed functions (\ref{vbg}) into the energy density function (\ref{e7}) and keeps  the terms with the  first order of  variations, it yields:
\begin{equation} \label{vare}
\delta\varepsilon(x,t)=\varphi^{'}(\delta\phi)^{'}+(\delta\phi)\frac{d U(\varphi,N_{2}\pi)}{d \varphi}.
\end{equation}
 Note that, for a non-moving $H$ ($V$)-soliton, it is obvious that  $\dot{\varphi}=0$  and $\frac{\partial U}{\partial \psi}$ ($\frac{\partial U}{\partial \phi}$) for $\psi=N_{2}\pi$ ($\phi=N_{1}\pi$) would be zero. Moreover, from Eq.~(\ref{e3})  for a non-moving $H$-soliton, it is easy to show that $\varphi^{''}=\frac{d U(\varphi,N_{2}\pi)}{d \varphi}$. Therefore, Eq. (\ref{vare}) is simplified   to
\begin{equation} \label{vare2}
\delta\varepsilon(x,t)=\varphi^{'}\delta\phi^{'}+(\delta\phi)\varphi^{''}=\frac{d}{dx}(\varphi^{'}\delta\phi)=\frac{d}{dx}(F).
\end{equation}
where $F=\varphi^{'}\delta\phi$. Since $\varphi^{'}$ is zero at $x=\pm \infty$, then the change  in the rest energy, to the first order of  variations, would be
\begin{equation} \label{vare2}
\delta E_{o}=\int_{-\infty}^{\infty}\delta\varepsilon(x,t)dx=F(\infty)-F(-\infty)=0.
\end{equation}
Thus, to the first order of  variations  $\delta E_{o}$ would be zero. If one considered  the second order of variations,   $\delta E_{o}$ is not zero anymore, but it is very small.  Hence,  the increase  of   the total rest energy for a small deformed  $H$ ($V$)-soliton (\ref{vbg}) is approximately  equal to zero, i.e. the rest energy of a little disturbed $H$ ($V$)-soliton is approximately equal to the  undisturbed one.

We know that $\phi$ and $\psi$ are both scalars, then  a  moving perturbed $H$-soliton should be written generally in the following form:
\begin{equation} \label{e24}
\phi_{v}(x,t)=\varphi(\gamma(x-vt))+N_{1}\pi,\quad\quad \psi_{v}(x,t)=N_{2}\pi+\Psi(\gamma(x-vt))e^{i(kx-\omega t+\theta)},
\end{equation}
where $kx-\omega t=k_{\mu}k^{\mu}$  is a scalar. If they  are  inserted  into  equation  (\ref{e4}) and expanding it to the first order in $\Psi$, we obtain
\begin{eqnarray} \label{e25}
&&(v^2\gamma^2\frac{d^2\Psi}{d\tilde{x}^2}+2iv\omega\gamma\frac{d\Psi}{d\tilde{x}}-\omega^2\Psi)e^{i(kx-\omega t+\theta)}-(\gamma^2\frac{d^2\Psi}{d\tilde{x}^2}+2ik\gamma\frac{d\Psi}{d\tilde{x}}-k^2\Psi)e^{i(kx-\omega t+\theta)} \nonumber\\&&=
(-2+2\alpha\sin^{2}(\varphi))\Psi e^{i(kx-\omega t+\theta)},
\end{eqnarray}
where $\tilde{x}=\gamma(x-vt)$. Omitting the common exponential factor from both sides and equating the real and imaginary
parts of the equation to zero separately, we obtain
\begin{equation} \label{e26}
k=\omega v,
\end{equation}
and
\begin{equation} \label{e27}
-\frac{d^{2}\Psi}{d\tilde{x}^{2}}+V(\tilde{x})\Psi=(\omega^2-k^2)\Psi=\omega_{o}^{2}\Psi.
\end{equation}
One can simply use  Eq.~(\ref{e26})  and the fact that $kx-\omega t=k_{\mu}x^{\mu}$ is a scalar to obtain
\begin{equation} \label{e28}
\omega=\gamma\omega_{o}.
\end{equation}
Moreover, it is easy to show that there is a similar relationship   for relativistic  energy of  soliton solutions, i.e. $E=\gamma E_{o}$. Therefore, frequency and energy have the same behavior and we can  relate them  via introducing a Planck-like constant $\overline{h}$:
\begin{equation} \label{e29}
E=\overline{h} \omega.
\end{equation}
Similarity, it is possible to find a  relation between  relativistic momentum of a soliton solution and wave number $k$:
\begin{equation} \label{e30}
p=\int_{-\infty}^{+\infty}T^{01}dx=m v=\overline{h} k.
\end{equation}
This equation  is very interesting, since it resembles the deBroglie's relation.

 $D$-solitons can also take an oscillating form after collisions.  For a $D$-soliton the related   dynamical equation and potential term were presented in Eqs.~(\ref{e31}) and (\ref{e32}) respectively.  In a similar way, we add a small time dependent oscillatory   term to the static solution as a new one:
\begin{equation} \label{e33}
\phi_{o}(x,t)=\chi(x)+\Phi(x)e^{-i(\omega_{o} t-\theta)},
\end{equation}
in which $\chi$ is the static kink (anti-kink) solution of Eq.~(\ref{e31}), i.e. the ones which are shown in the Fig.~(\ref{Dsolitons}). We can now substitute this ansatz (\ref{e33}) in the wave equation (\ref{e31}) and expand the potential by  keeping only the linear terms in $\Phi$. Finally,  the  same Schr\"{o}dinger-like eigenvalue equation  is obtained again
\begin{equation} \label{e36}
-\frac{d^{2}\Phi}{dx^{2}}+V(x)\Phi=\omega_{o}^{2}\Phi,
\end{equation}
where
\begin{equation} \label{e38}
V(x)= \dfrac{d^2 {\cal U}(\chi)}{dx^2}
\end{equation}
and ${\cal U}(\chi)=\sin^{2}(\chi)-\frac{\alpha}{2}\sin^{4}(\chi)$. For such systems, again there is a trivial solution $\Phi=\xi \dfrac{d\chi}{dx}$ with  $\omega_{o}=0$    \cite{DC,G,Riazi4,OV,GH,MM1}.  Moreover, it can be shown numerically that for systems with $0<\alpha<2$, there is a  non-trivial internal mode which can be interpret as a physical  channel for absorbing  fluctuations and  generates  permanent oscillations in $D$-solitons. Finally, for a moving perturbed $D$-soliton, the corresponding results are exactly the same as those obtained for a  perturbed $H$ and $V$-solitons.

The fore-mentioned results reveal further the interesting wave-particle aspects of solitons. Moreover, we will show numerically  in the next section how the wave aspect can cause  an uncertainty in the outcome of collisions.

\section{UNCERTAINTY IN COLLISIONS}\label{sec6}

\begin{figure}[ht!]
   \centering
   \includegraphics[width=130mm]{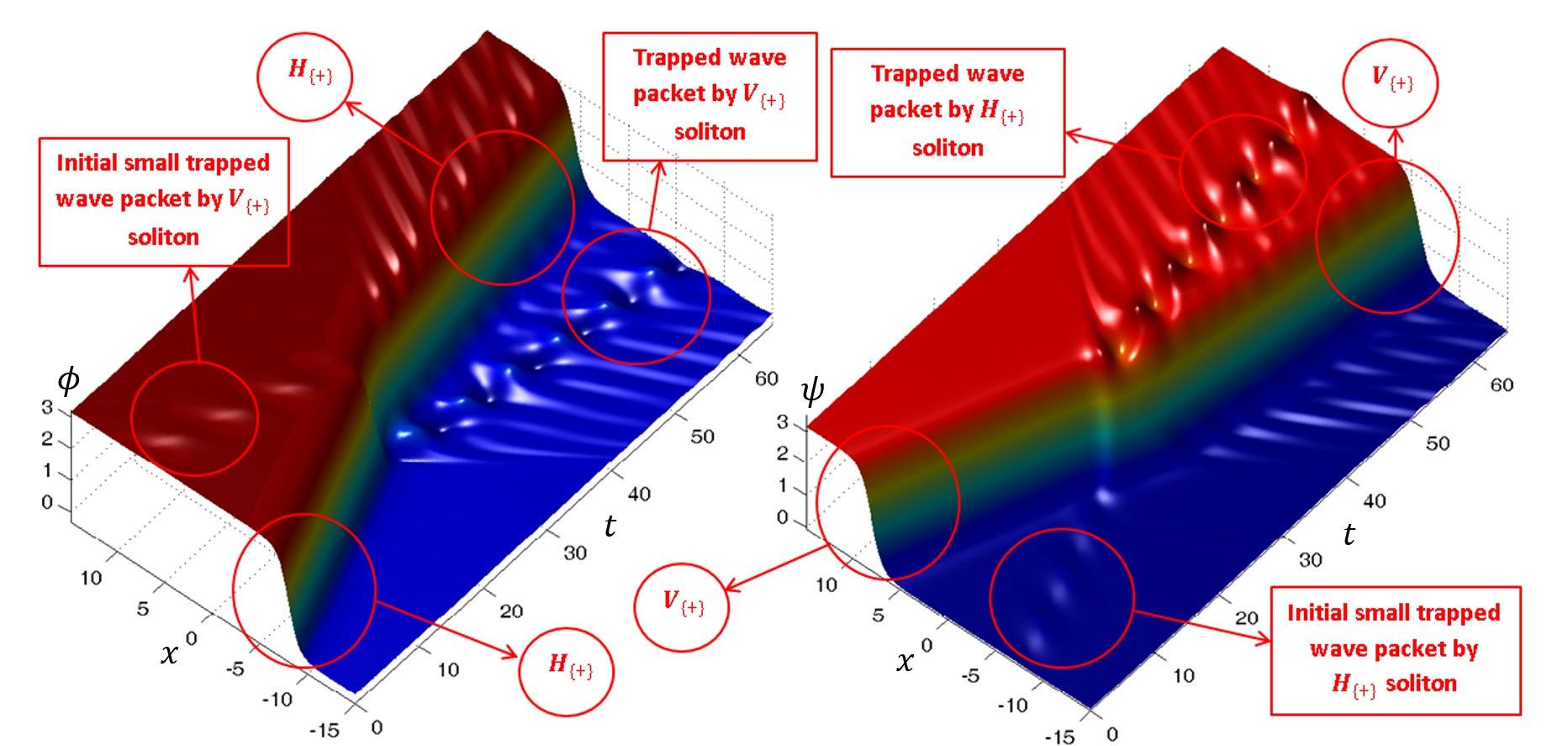}
   \caption{ $\phi$ and $\psi$ representation of a disturbed $H_{\{+\}}-V_{\{+\}}$ collision with $0.3$ initial speed for which $\alpha=1.2$ and $b=-a=8$. The initial phases of the small trapped wave packets are  $\theta_{1}=0$ and $\theta_{2}=0$. At $t=0$ the initial small trapped wave packet by $H_{\{+\}}$ ($V_{\{+\}}$) is the same oscillatory   term $\Psi\cos{(-kx-\omega t+\theta_{2})}$ ($\Phi\cos{(kx-\Omega t+\theta_{1})}$) in Eq.~(\ref{e37}).} \label{DisHV2}
  \end{figure}

If a small wave packet can be  absorbed by a soliton, we will have an disturbed  soliton. To study a disturbed $H-V$ collision with initial speed $v$, for which at least one of  the $H$ and $V$-solitons get excited, we first need to prepare proper initial conditions for $\phi$ and $\psi$ fields in the following form
\begin{eqnarray} \label{e37}
&&\phi_{v}(x,t)=\varphi(\gamma[(x-a)-vt])+\Phi(\gamma[(x-b)+vt])\cos{(kx-\Omega t+\theta_{1})},\nonumber\\&&
\psi_{v}(x,t)=\varphi(\gamma[(x-b)+vt])+\Psi(\gamma[(x-a)-vt])\cos{(-kx-\omega t+\theta_{2})},
\end{eqnarray}
where $a$ and $b$ are  initial positions provided $b-a$ is large enough and $\theta_{1}$ and $\theta_{2}$ can be some arbitrary initial phases. The small eigenfunctions $\Phi$ and $\Psi$, and eigenfrequency $\Omega=\omega=\gamma\omega_{o}=\gamma\Omega_{o}$,  can be obtained easily by a straightforward Runge-Kutta method for Eq.~(\ref{e22}).
One can see Fig.~\ref{DisHV2} at $t=0$ for better understanding.

\begin{figure}[ht!]
  \centering
  \includegraphics[width=170mm]{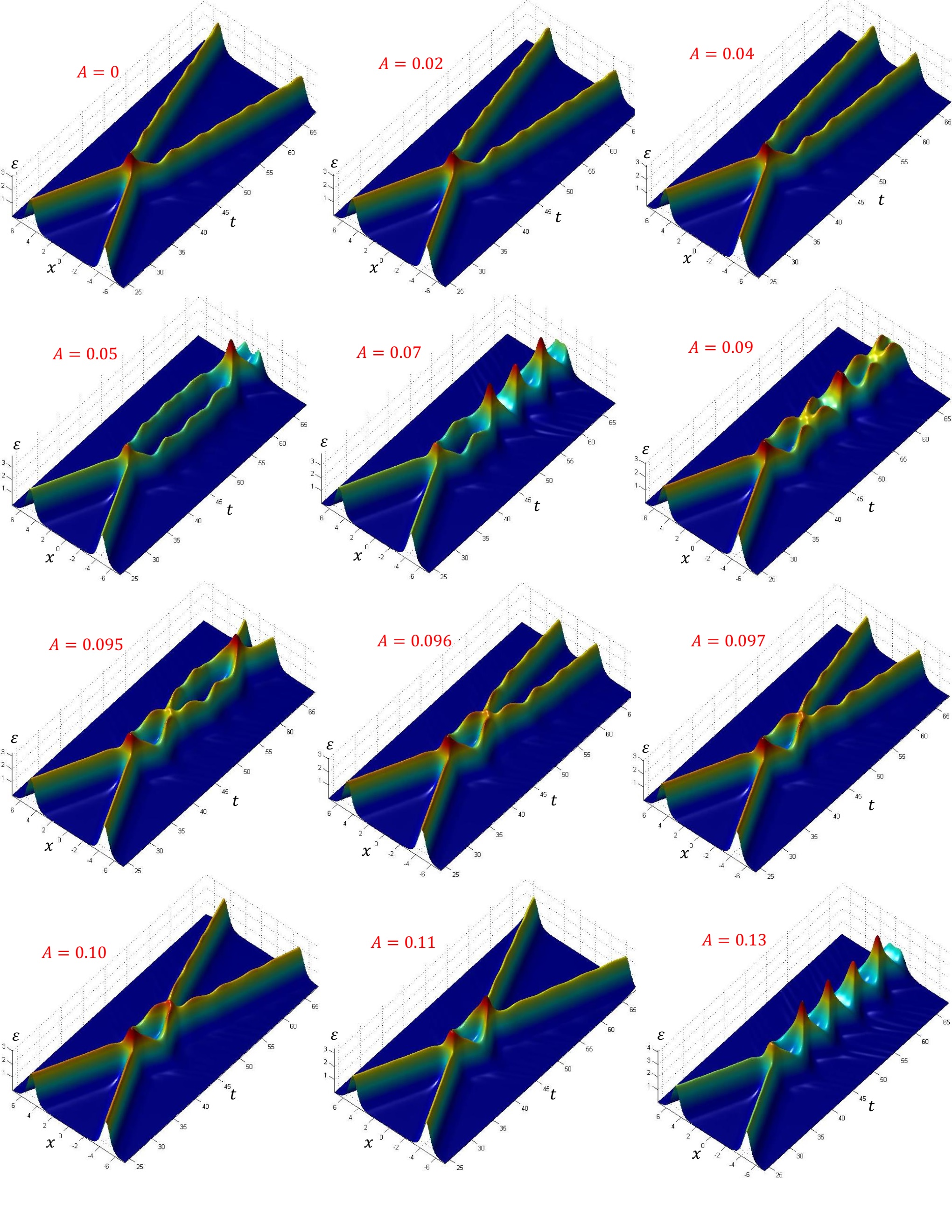}
  \caption{Energy density representations of a disturbed $H_{\{+\}}-V_{\{-\}}$ collision. We have set  $v=0.3$, $\alpha=1$, $a=-11.9$, $b=11.9$ and $\theta_{1}=\theta_{2}=0$. They  show  that  different maximum  amplitude of the initial small trapped wave packets  lead to different fates of the collision.} \label{amp}
 \end{figure}

The small wave packets  does not seem to cause  a considerable  change in the particle aspect of a soliton.  We consider wave packets which lead to  less than $0.3$ percent change in the soliton energy.  It was observed  numerically that these small wave packets can have a considerable  affect on the collision fate.
For example, for an arbitrary  system with   $\alpha=1$, if one considers  a disturbed  pair  $H_{\{+\}}$-$V_{\{-\}}$   which both initially  trap small  wave-packets with the same   maximum amplitude  ($A=\textrm{max}(|\Phi|)=\textrm{max}(|\Psi|)$) and initial  phases  $\theta_{1}=\theta_{2}=0$, when they are initialized to be at $a=-b=-11.9$ with $v=0.3$, depending  on different values for  the maximum  amplitude $A$, there are different fates for the collision (see Fig.~\ref{amp}).

\begin{figure}[ht!]
  \centering
  \includegraphics[width=130mm]{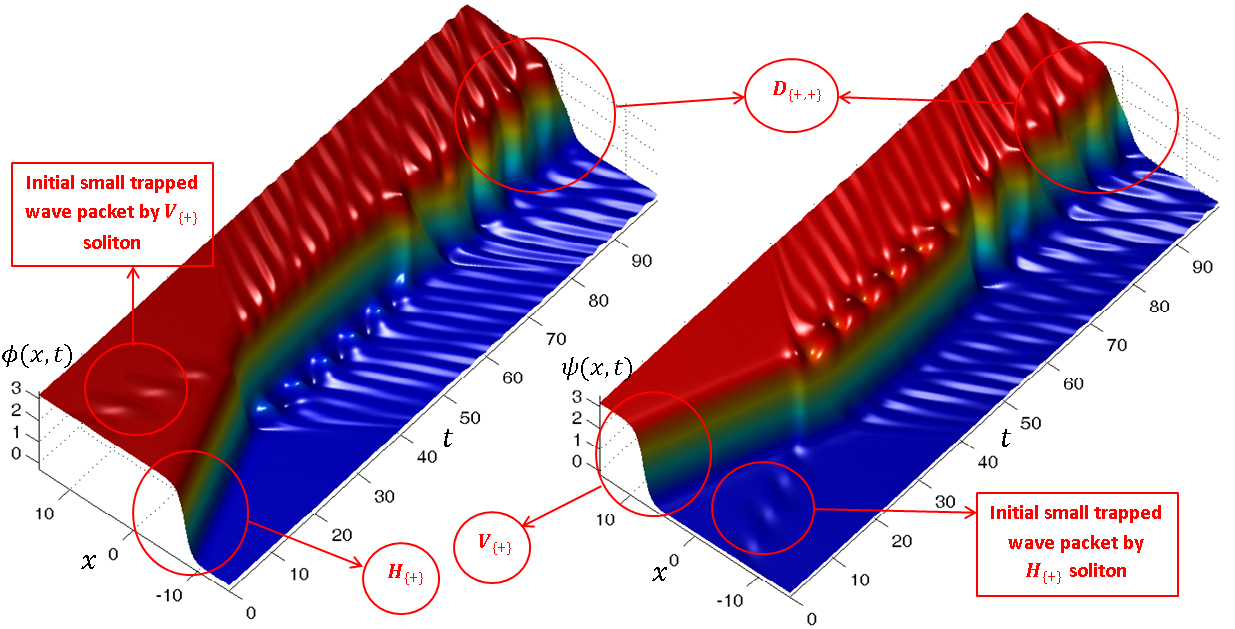}
  \caption{ $\phi$ and $\psi$ representation of a disturbed $H_{\{+\}}-V_{\{+\}}$ collision with $0.3$ initial speed for which $\alpha=1.2$ and $a=-b=8$. The initial phases of the trapped wave packets are  $\theta_{1}=0$ and $\theta_{2}=\frac{\pi}{2}$. } \label{DisHV}
 \end{figure}

\begin{figure}[ht!]
    \centering
    \includegraphics[width=150mm]{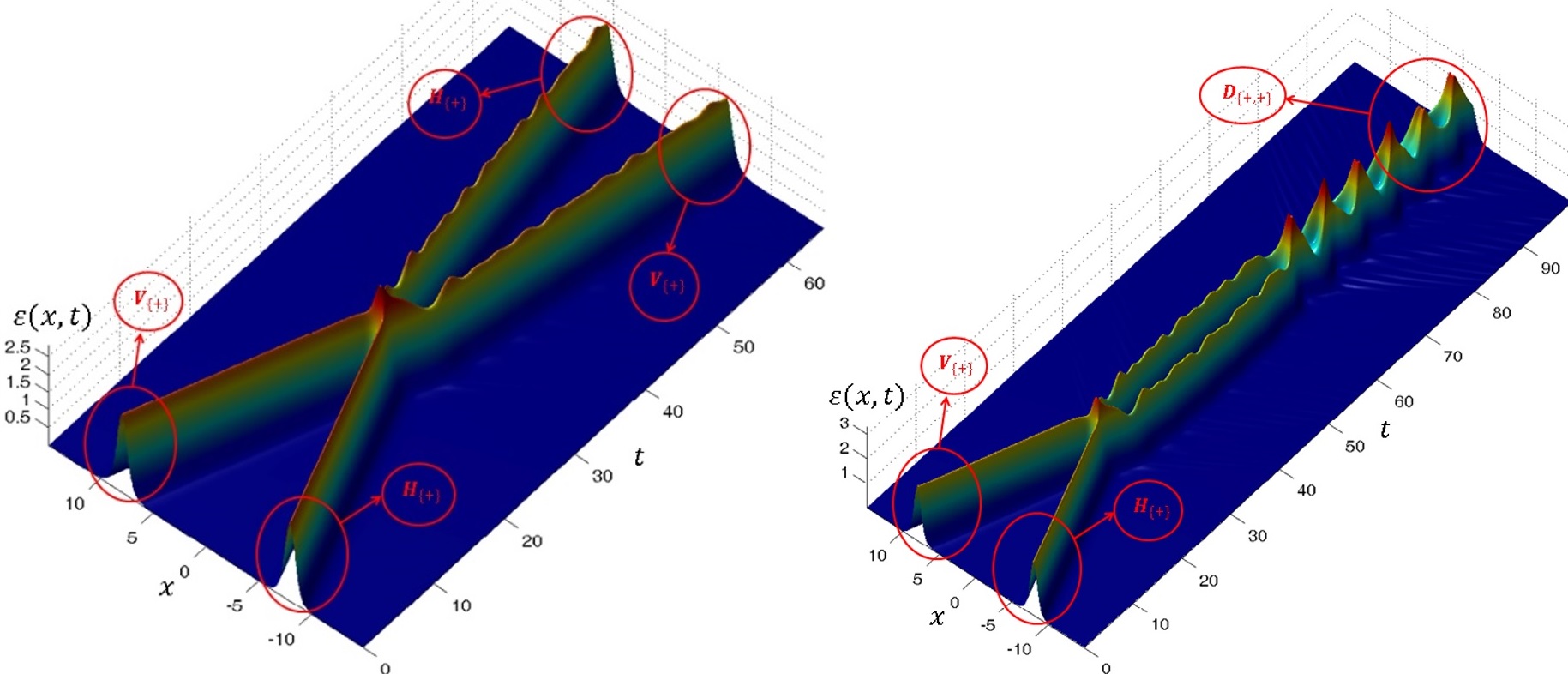}
    \caption{Energy density representation of a disturbed $H_{\{+\}}-V_{\{+\}}$ collision with $0.3$ initial speed for which $\alpha=1.2$ and $a=-b=8$. The initial phases for  the trapped wave packets of the left (right) Fig are  $\theta_{1}=0$ and $\theta_{2}=\frac{\pi}{2}$ ($\theta_{1}=0$ and $\theta_{2}=\frac{\pi}{2}$). } \label{DDD}
   \end{figure}

One might think that  $\theta$  as an optional initial phase is an unimportant parameter. In fact,
it has no role in  determining  basic physical features of a single soliton such as  energy, momentum, topological charge and  eigenvalues $\omega_{o}^{2}$.  But, it was seen numerically during the collision  between disturbed solitons, these initial phases   become very important.
For example, similar to  Fig.~\ref{ghablphi} as an undisturbed $H_{\{+\}}$-$V_{\{+\}}$ collision, one can study  a new disturbed  version  for which the   amplitudes  of the initial small  trapped wave packets be equal to $A=0.04$. Hence,  if one sets $a=-b=-8$, $\alpha=1.2$ and $v=0.3$,  it was shown numerically that  related to  different initial phases, different fates will happen. Namely, for $\theta_{1} =\theta_{2}= 0$,  solitons are scattered from each other and reappear after collision, but with a periodic oscillations in the amplitude of their energy density functions (see Fig.~\ref{DisHV2} and the left hand side of Fig.~\ref{DDD}). If $\theta_{1} =0 $ and $\theta_{2}= \frac{\pi}{2}$, they capture each other and make a disturbed  D-soliton (see Fig.~\ref{DisHV} and the right hand side of Fig.~\ref{DDD}).

\begin{figure}[ht!]
  \centering
  \includegraphics[width=170mm]{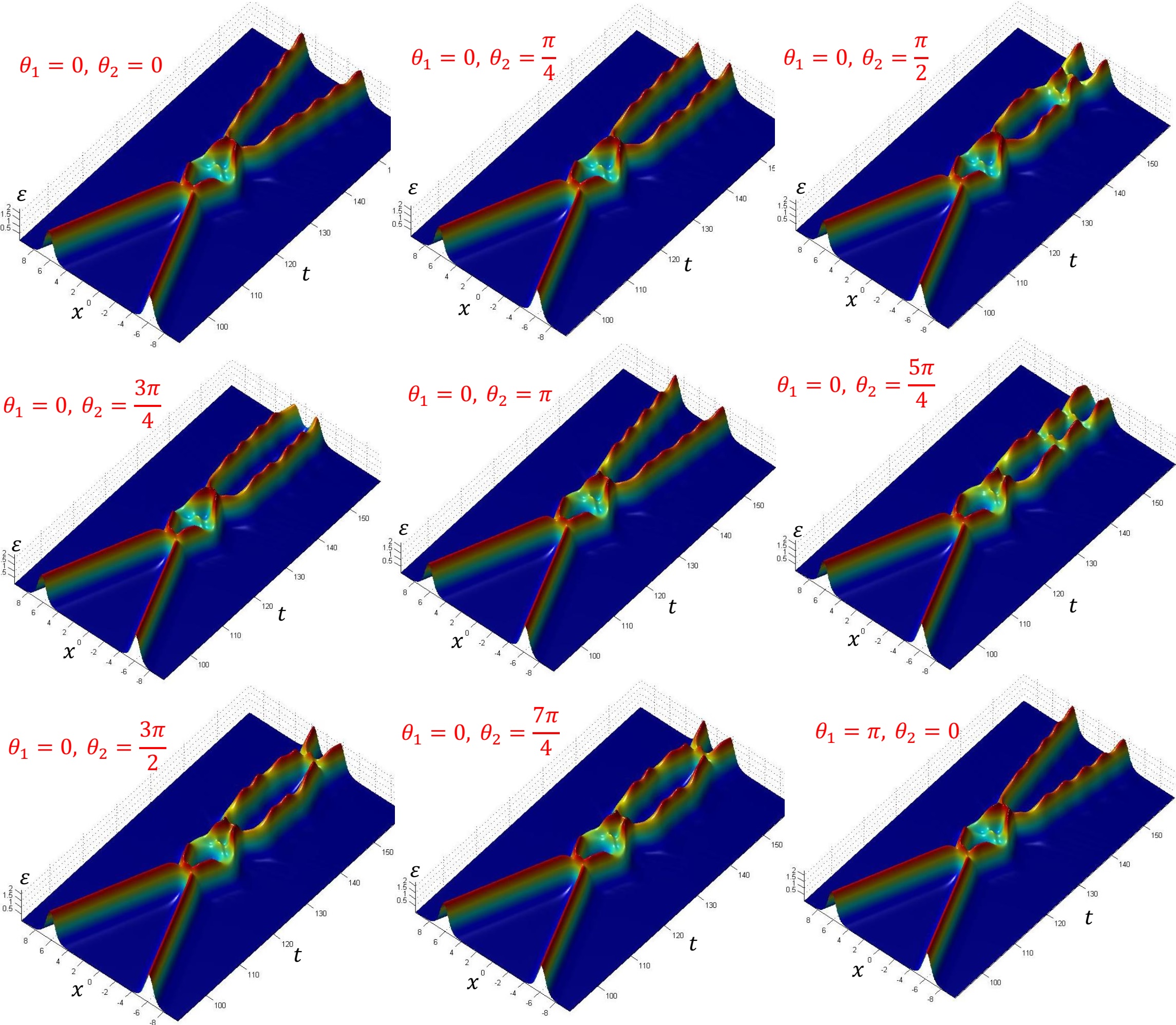}
  \caption{Energy density representations of a disturbed $H_{\{+\}}-V_{\{-\}}$ collision. We have set  $v=0.2$, $\alpha=1.5$, $a=-24.5$, $b=24.5$ and $A=0.05$.  They  show  that different  initial phases  of the trapped wave-packets lead to different fates of the collision.} \label{phase}
 \end{figure}

In a more complete example, to show numerically how such initial phases lead to different fates in a collision process, we  can study many disturbed  $H_{\{+\}}-V_{\{-\}}$ collisions  with  different  arbitrary initial phases of a   system with $\alpha=1.7$ for which $a=-24.5$, $b=24.5$, $A=0.05$ and $v=0.2$.  Now,   without  any change in the physical features of the initial functions (\ref{e37}),  we are free to choose  any value for the initial phases $\theta_{1}$ and $\theta_{2}$. The different  results of the collision   for different initial phases can be seen in  Fig.~\ref{phase}. Again, it demonstrates  that different initial phases lead to different fates of the collision.
Then, both capture and scattering can occur depending on the initial phases $\theta_{1}$ and $\theta_{2}$. It is reasonable to expect  that if the amplitude of the initial tapped wave packets tend to zero  the influence of initial phases on the collisions tends to zero as well.

\begin{figure}[ht!]
  \centering
  \includegraphics[width=130mm]{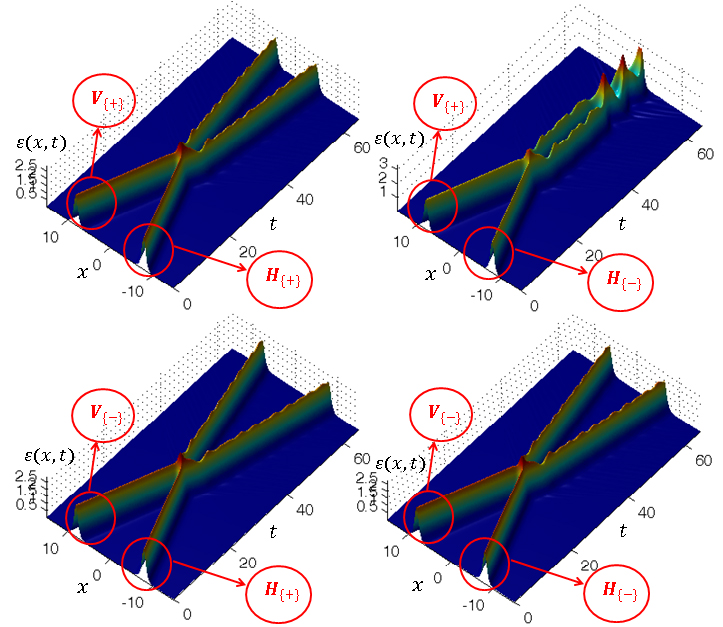}
  \caption{Energy density representations for different type of a disturbed $H-V$ collision with $0.3$ initial speed when $\alpha=1.2$. The initial phases of the trapped wave packets are  $\theta_{1}=-\frac{\pi}{3}$ and $\theta_{2}=0$. The same amplitude of the initial small trapped wave packets by $H$ and $V$-solitons have set  to  $A=0.04$. } \label{4E}
 \end{figure}

Moreover, numerically, one can show that there is no  difference  between undisturbed  $H_{\{+\}}-V_{\{+\}}$,  $H_{\{+\}}-V_{\{-\}}$,  $H_{\{-\}}-V_{\{+\}}$ and  $H_{\{-\}}-V_{\{-\}}$ collisions in the energy density representation,  i.e. all of them will give the same energy density figures after collisions.  But  for  disturbed ones, it was seen numerically that depending on what type of $H$ and $V$ to use, the resulting energy density figures will be different  (see Fig~\ref{4E}).

\begin{figure}[ht!]
  \centering
  \includegraphics[width=170mm]{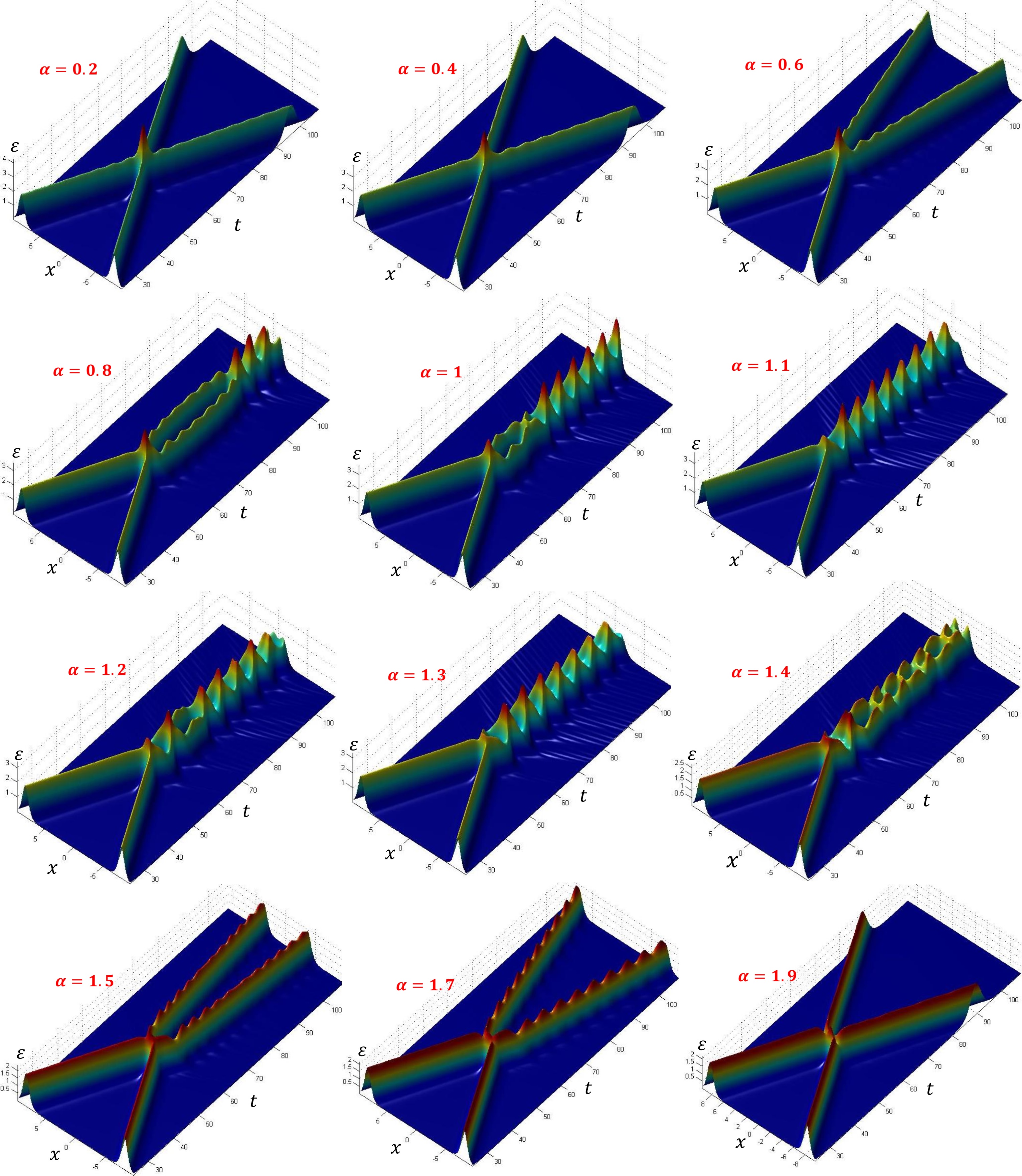}
  \caption{Energy density representations  of a  disturbed $H_{\{+\}}-V_{\{-\}}$ collision for different $\alpha$'s. We have set $\theta_{1}=0$, $\theta_{2}=0$, $v=0.25$, $b=-a=14$ and $A=0.05$.} \label{Dalpha}
 \end{figure}

\begin{figure}[ht!]
  \centering
  \includegraphics[width=170mm]{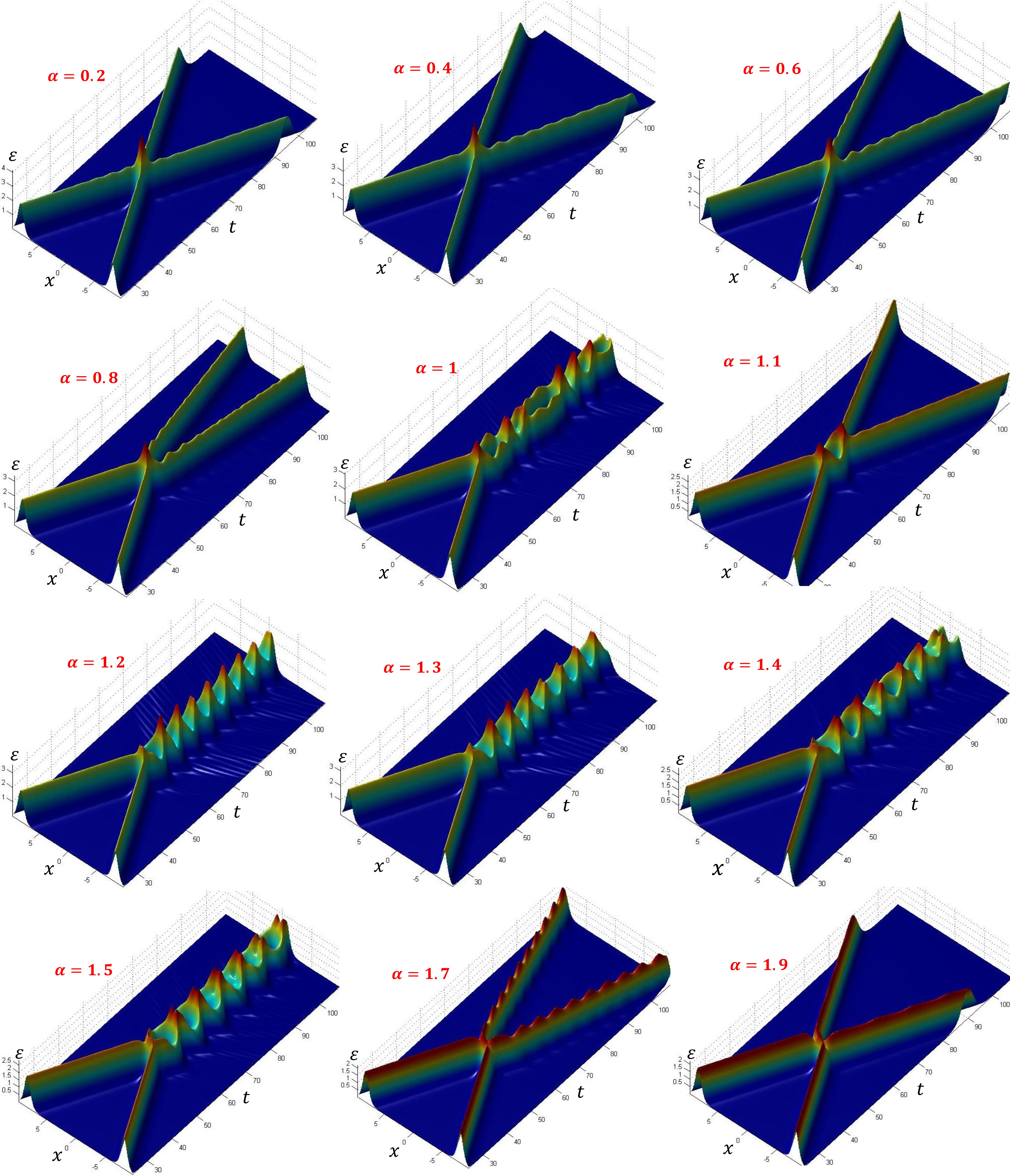}
  \caption{Energy density representations  of a  disturbed $H_{\{+\}}-V_{\{-\}}$ collision for different $\alpha$'s. We have set $\theta_{1}=0$, $\theta_{2}=\pi$, $v=0.25$, $b=-a=14$ and $A=0.05$.} \label{Dalpha2}
 \end{figure}

\begin{figure}[ht!]
  \centering
  \includegraphics[width=170mm]{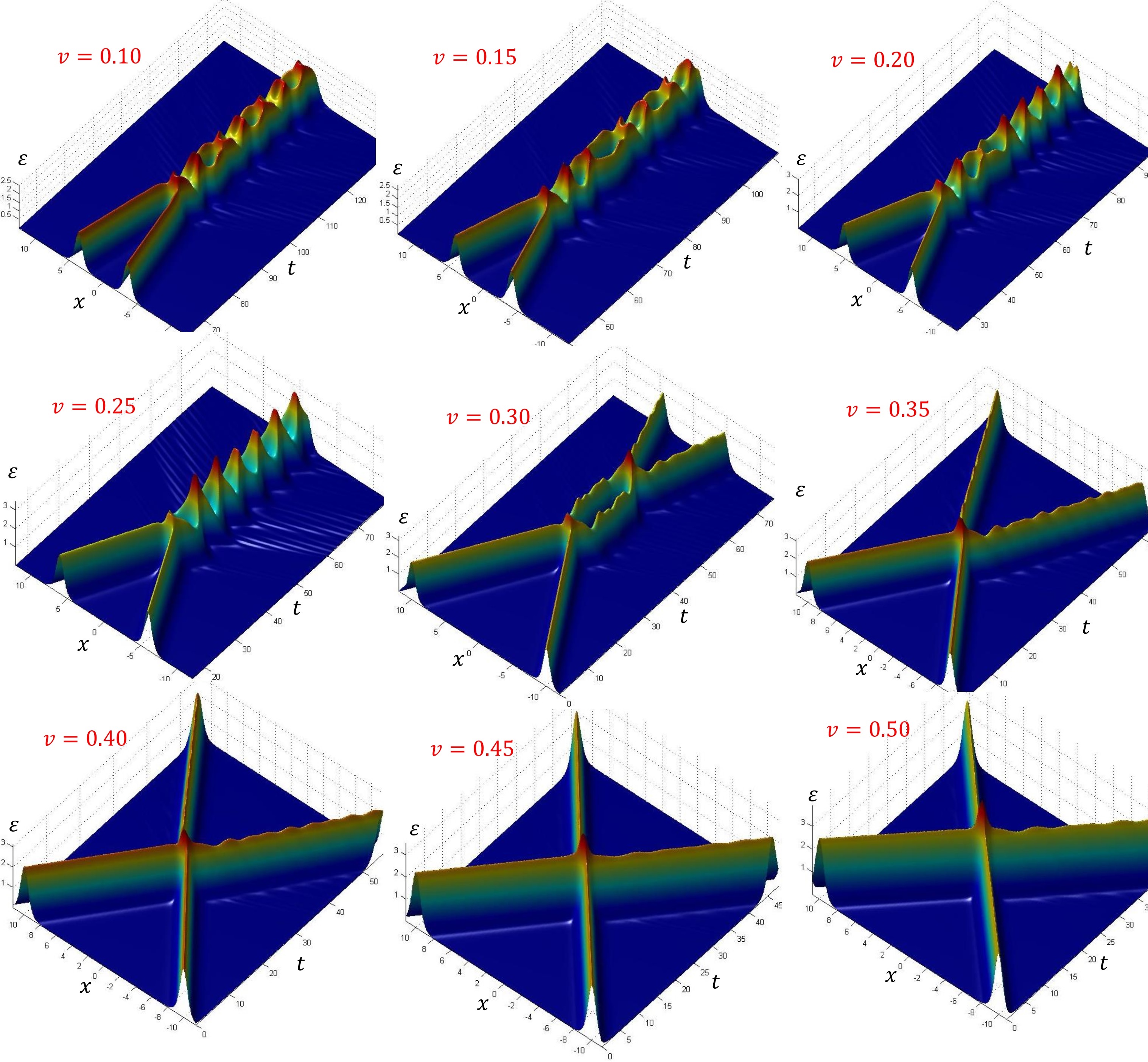}
  \caption{Energy density representations  of a  disturbed $H_{\{+\}}-V_{\{-\}}$ collision for different velocities. We have set $\alpha=1.2$, $\theta_{1}=0$ and $\theta_{2}=0$,   $a=-b=-10$ and $A=0.05$.} \label{vlo}
 \end{figure}

\begin{figure}[ht!]
  \centering
  \includegraphics[width=170mm]{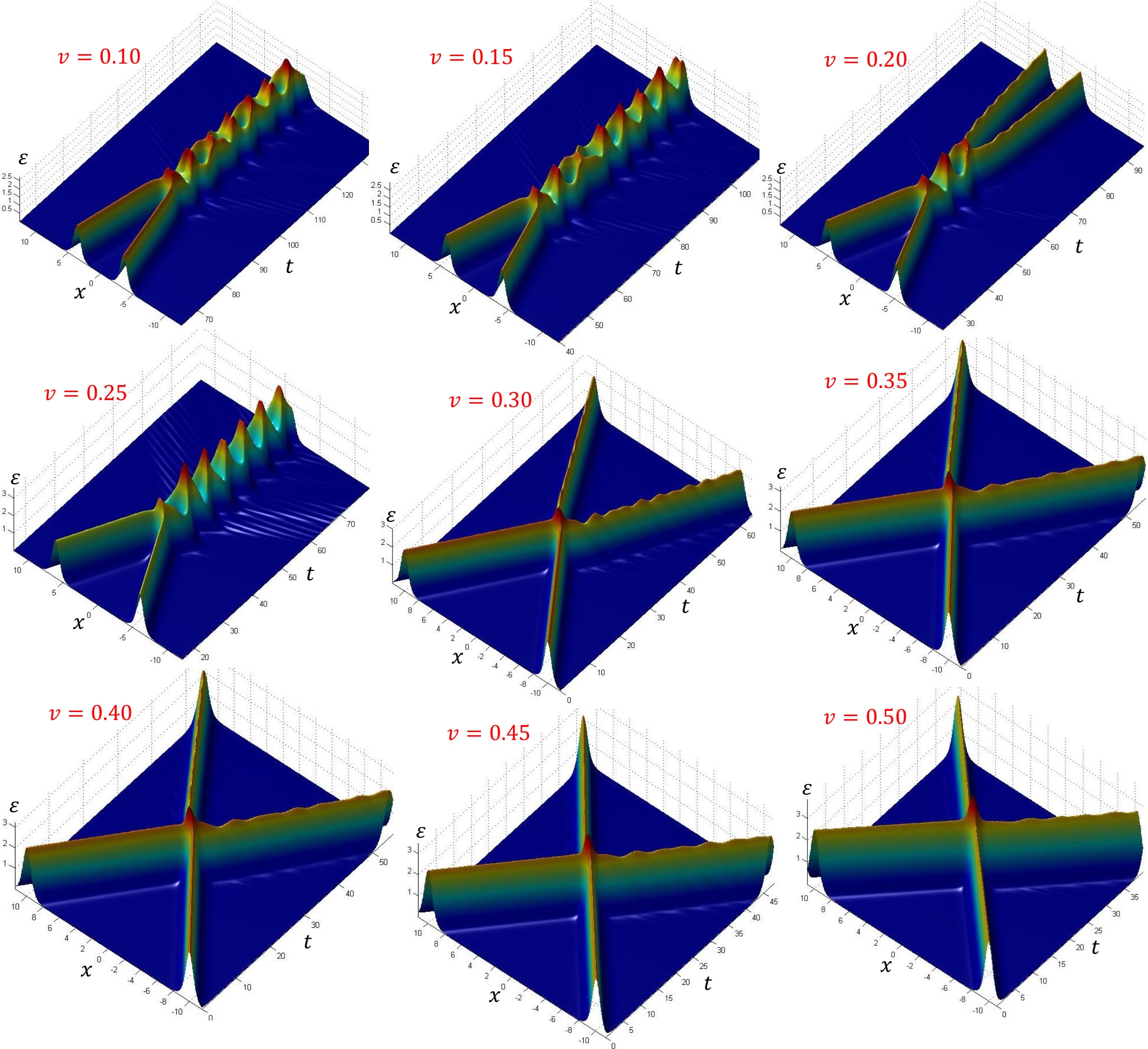}
  \caption{Energy density representations  of a  disturbed $H_{\{+\}}-V_{\{-\}}$ collision for different velocities. We have set $\alpha=1.2$, $\theta_{1}=0$ and $\theta_{2}=\pi$,   $a=-b=-10$ and $A=0.05$.} \label{vlo2}
 \end{figure}

To more support, two  different situations for disturbed $H_{\{+\}}-V_{\{-\}}$ collisions  can be  studied. First, for fixing initial phases  $\theta_{1}=0$, $\theta_{2}=0$ ($\theta_{2}=\pi$), initial speed $v=0.25$, the same maximum  amplitude  of the small trapped wave packets $A=0.05$,  and  initial positions $b=-b=14$,  then we can study the collision fates for  different $\alpha$'s (see Fig.~\ref{Dalpha} and Fig.~\ref{Dalpha2}).   Second,  for fixing $\theta_{1}=0$, $\theta_{2}=0$,   $A=0.05$, $\alpha=1.2$, and   $a=-b=-10$,  then we can study the collision fates for  different velocities (Fig.~\ref{vlo} and Fig.~\ref{vlo2}). It is seen numerically that the high speed  collisions   (energetic collisions) reduce the influence of the initial trapped wave packets in the collision  fates, i.e. we do not see significant different  fates   in the outputs of the  energetic collisions (compare Fig.~\ref{vlo} with  Fig.~\ref{vlo2} for collisions with initial  speeds larger than $v=0.4$). Moreover, comparting Fig.~\ref{Dalpha} with Fig.~\ref{Dalpha2} reveals that the systems for which $\alpha$ are close to 0 or 2,  are less affected by the initial trapped wave packets properties.

  Therefore, there is an  apparent  uncertainty    in the collision process  which  originates from the amplitudes and  initial phases of the  trapped perturbations. The initial phase does not  play any crucial role in the particle aspect of a single soliton, but it is the main reason  of the uncertainty in collisions, i.e.  it can lead to completely different fates.   However,  for undisturbed solitons there is no uncertainty and everything is   predictable. Of course, in real word, random phase perturbation are inevitable, and therefore we might expect probabilistic behaviour in soliton scattering. In fact, the initial phases, in this model, behave like a hidden variable  which   cause to a probabilistic behaviour for final fates of different particle-like solutions in collision processes.

In this paper, we showed that in general,   as  seen in the figures,  which  two solitons capture each other  always lead to similar  breathing oscillating structures after collisions. However, these  breathing oscillating structures have been seen  in  many different nonlinear systems in $1+1$ and $2+1$ dimensions similarly \cite{G,R1,R2,MM2}.  We think that this similarity, just is referred to the non-linearity nature of them. That is, apparently, the emergence of such structures is a common feature   that is related to the nature of the nonlinear systems. But, in general there is not  any other meaningful relationship between them. For example,  the special cases which were seen in  \cite{R2} are happened when some special conditions are fulfilled, while there are not such restrictive conditions  in our model and other  kink-bearing models to see such breathing oscillating structures. Moreover, the kink-bearing systems  are essentially relativistic, while the other systems \cite{R1,R2}  are  non-relativistic and this is another difference.

\section{SUMMERY AND CONCLUSIONS}\label{sec7}

Inspired by the well-known  sine-Gordon (SG) system, we presented a coupled system of the non-linear PDEs  which was built
by two real   scalar fields $\phi$ and $\psi$  in $1+1$ dimensions. A free parameter $\alpha$ was introduced to characterize  the coupling provided $\alpha\leqslant 2$. For $\alpha=0$ and $\alpha=2$, we retrieved two independent SG systems.    The coupled system was shown to have three different $H$, $V$  and $D$-soliton solutions.  $H$ and $V$  solutions are nothing but the  known SG kink (anti-kink) solutions for fields $\phi$ and $\psi$ respectively, and $D$ solutions are a composite of $H$ and $V$ solions.  It was seen numerically that  for systems with $0<\alpha<2$, there are always  some permanent small wave packets which are trapped inside solitons after collisions.  We  classified  $H$, $V$ and $D$-solitons with  introducing subscripts $\{+\}$ and $\{-\}$ to recognize their topological charges.

It was seen analytically that for systems with $0<\alpha<2$, there is an internal mode for  $H$, $V$ and $D$-solitons which makes  them  able to trap a small wave packet.  A wave packet is characterized by an  eigenfunction and a special frequency $\omega_{o}$
(or eigenvalue  $\omega_{o}^{2}$) of a Schr\"{o}dinger-like equation.  We found numerically that there is an uncertainty in collision fates between disturbed solitons, i.e. the ones which initially trap  small wave packets. This uncertainty originates from the amount of the maximum  amplitudes and  initial   phases of the  small trapped wave packets by solitons.
For different initial phases, particle aspect of the solitons remain unchanged, while the final behaviour may be drastically affected. Nevertheless, the energetic collisions reduce  the influence of the initial trapped wave packets on the  collision fates. Moreover, the systems for which $\alpha$ are close to 1 or 2  are less affected by the initial trapped wave packets properties.

 \section*{ACKNOWLEDGEMENT}

The authors acknowledge the Persian Gulf University Research Council.

\end{document}